\documentclass[aps,pre,nofootinbib,amsmath,amsfonts]{revtex4}
\usepackage{hyperref}
\usepackage{graphicx}

\newcommand{\f}{\frac}

\graphicspath{{figuresmu/}}
\begin{document}
\title{$\mu$-model for the statics of dry granular medium}
\author{K.~S.~Glavatskiy}
\email{k\_glavatskiy@ukr.net} \affiliation{Department of
Chemistry, Norwegian University of Science and Technology,
Trondheim, Norway}
\author{V.~L.~Kulinskii}
\affiliation{Department of Theoretical Physics, Odessa National
University, Dvoryanskaya 2, 65026 Odessa, Ukraine}
\begin{abstract}
We propose the description of the granular matter which is based
on distribution of dry friction coefficients. Using such a concept
and a simple one-dimensional packing of grains we solve the silo
problem. The friction coefficients at contacts are determined both
by geometry of packing configuration and the stress distribution
in a medium. Within such an approach the Janssen coefficient
$k_{J} $ is determined and its dependence on the particle-particle
and boundary-particle friction coefficients is obtained. Also we
investigate the conditions for the appearance of the maximum in
the pressure distribution with the depth with overweight on top.
As an outcome of our work we propose the general framework to the
description of the granular matter as a continual medium which is
characterized by the field of the dry friction tensor.
\end{abstract}
\maketitle 

\section*{Introduction}
The properties of granular matter differ drastically from those of
other continuous media like solids, liquids and gases. Despite its
mechanical nature the problem of description of granular media is
still open problem. Some arguments put forward recently
\cite{du95,kadanoff/rmp/1999} even question the possibility of the
description of such media basing on the hydrodynamic approach. The
latter usually applied to continuous media with short-range
interparticle forces. But nonconservative nature of the friction
force hampers direct application of such an approach based on
local balance equations for physical quantities like momentum,
energy, entropy, etc. In such a complex situation with the
dynamics the static properties of the granular medium are easier
to investigate. There are several characteristic static phenomena,
which any theory should explain. They are connected with the
static distribution of the pressure along a silo. Namely, the
deviation from Pascal law and the nonmonotonic dependence of the
apparent mass on the overweight on the top of a silo. Lets us
review them shortly.

In contrast with conventional fluids, "hydrostatic" pressure in
granular media reaches the asymptotic finite value $p_0$ at some
finite depth $\lambda$, and can be described as
\begin{equation}\label{janss}
p(z) = p_0 \left(1-\exp(-z/\lambda)\right)\,,\quad \lambda =
\f{R}{2\mu k_J}
\end{equation}
where $R$ is the width of the silo, $\mu$ is the static friction
coefficient between the grains and the walls of the silo, and
$k_J$ is the so-called Janssen coefficient. First explanation of
this behavior originated from the Jannsen model \cite{janssen}
(see also \cite{degennes/rmp/1999}), based on two simple
assumptions: i) the linear relation between $\sigma _{xz} $ and
$\sigma_{xx} $ components of the stress tensor, which is analogous
to the Coulomb-Amonton law for the dry friction, and ii) the
linear ratio between other two components of the stress tensor:
$\sigma _{xx} = k_{J} \sigma _{zz}$. Since then, many approaches
to the statics of granular medium have been proposed either giving
the grounds for the Janssen relations
\cite{ovralezclement/pre/2003,clement/epj/2005}, or building the
theory without using them
\cite{pitman/pre/1998,socolar/pre/1998,coppersmith/pre/1996}.

One of the simplest approach treats the granular media within the
framework of linear isotropic elasticity. Indeed, the Janssen
coefficient $k_{J}$ might be expressed through the Poisson ratio
of the media, and numerics recovers the Janssen relations between
the components of the stress tensor \cite{clement/epj/2005}.
Detailed comparisons with experimental results even allow to
extract the relation between the Poisson ratio and the granular
packing fraction \cite{clement/epj/2005}. However, such a picture
is hardly adequate for the granular media of rigid particles. In
addition, it is hard to control realization of the Coulomb
threshold condition everywhere at the walls, which, obviously,
influences the experimental results \cite{vanel/inbook/98}.

Other approaches do not exploit elastic nature of the grains, but
use the Coulomb-Amonton law for the stress tensor components
inside the media or at the walls \cite{cates/prl/2000}. To compete
with indeterminancy of the problem, the granular media is assumed
to be at the verge of Coulombic failure everywhere in the bulk.
However, this assumption about fully mobilized friction seems to
be too restrictive \cite{degennes/rmp/1999}, since the frictional
forces can be varied provided that the medium stays at rest. In
order to simplify the problem which include randomness in the
distribution of such coefficients, specific assumptions about the
geometrical distribution of the grains should be made, e.g. within
some lattice models \cite{coppersmith/pre/1996, socolar/pre/1998}.
Such models give the results which are consistent with continuum
theories for the average stresses.

A particularly interesting issue is the weight distribution in a
silo with some overweight on top. Experiments show the maximum in
the pressure distribution with the depth, in contradiction with
predictions of the simple Janssen model \cite{cates/prl/2000}.
Interestingly, this maximum appears almost at the same depth where
pressure at the silo without the overweight changes in $e$ times.
Although some of mentioned approaches do not contradict
experiment, but the physical reasons of this phenomenon are still
unclear.

The aim of this paper is to propose the description of the
granular media which incorporates an additional dry friction
tensor field. In the simplest 1D geometry considered below it
reduces to the distribution of the dry friction coefficients. We
show that in this case it is possible to get the expression for
the Janssen coefficient. Besides, we are able to explain the the
weight distributions in experiments with overweight as a result of
the inhomogeneity in the friction coefficient distribution. The
existence of characteristic scale for such inhomogeneity leads to
the parametric dependence of the relation ``apparent mass - filled
mass`` on  the ratio of two length scales characteristic
inhomogeneity and the saturation length for the bulk pressure.

The structure of the paper is as follows. In
Section~\ref{section1} we introduce the $\mu $-model and
illustrate it on the simple 1D granular packing, which resembles
the real granular packing in silo. The results and comparisons are
presented in Section~ \ref{section2} of the paper. We discuss the
force distribution with and without overweight for simple packing
and also find how Janssen coefficient $k_{J}$ can be evaluated
from microscopic characteristics of the media. In
Section~\ref{section3} we discuss the grounds of $\mu$-model using
the results in Section~\ref{section2}. Conclusions are given in
Section~\ref{conclusion}.

\section{$\mu$-model for simple 1D packing}\label{section1}
The grains at contact are subjected to static friction force. The
value of such a force is determined both by the geometry of the
contact surfaces and the stress at the contact. The value of the
friction force is between 0 and the maximum value. The last is
described by the Coulomb-Amonton law for static friction. Exact
value is defined by the condition to keep the grain at rest. Since
a grain is in contact with its neighbors at several points the net
force is the sum of reaction forces for each contact. Because of
this, granular system cannot be described using deterministic
approach. There are more unknown variables than the equations. The
granular matter can implement different force configurations even
if the packing configuration is the same.

Between two grains the reaction force acts. It is expedient to
split it into two components: normal (normal reaction force) and
tangential (friction forces). For each point of contact we can
write
\begin{equation}
\label{eq1} F_{\tau}  = \mu \cdot F_{n}\,,
\end{equation}
where coefficient $\mu $ is different for different contact point
but it's value is in the interval $[0,\mu _{f} ]$, where $\mu _{f}
$ is the static friction coefficient and determines the static
friction angle. Obviously these coefficients for each contact
point together with the normal stress determine the force
configuration of the system. In the continuous limit, the set of
these coefficients transforms into the tensor field $\mu $. Such a
tensor field becomes additional characteristic of a granular
medium.

\begin{figure}[hbt!]
\centering
\includegraphics[scale=0.3]{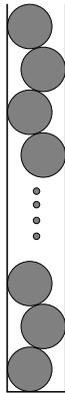}
\caption{Pseudo one-dimensional packing of identical
spheres}\label{fig1}
\end{figure}

To illustrate the idea we apply this model to simple granular
packing shown in Fig.~\ref{fig1}. It is a regular pseudo-1D
packing of identical rigid spheres in the vertical chute with a
proper width. We choose 1-D model in order get explicit analytical
results. Though such models seem to be oversimplified they grasp
the main feature of the granular systems, namely the jamming. Also
they are widely used in modelling avalanches within framework of
SOC \cite{soc}. In particular as is shown in \cite{du95} 1-D
systems shows nontrivial dynamic which is due to nonpotential
character of the interparticle interactions. Moreover, according
to \eqref{eq1} the coefficient of friction $\mu$ is defined in a
way irrespective on the dimension of the system. Thus it can be
defined both for the grains and for the mesoscopic regions of the
granular matter thus giving rise to the introduction of the
$\mu$-field.

Denote the normal force between grains $i$ and $i+1$ by $P_{i,i +
1}$, and between the grain $i$ and the wall by $N_{i}$. Let us
denote the friction coefficients for each ``grain-grain'' pair by
$\mu _{i,i + 1} $, and the ones for the ``grain-wall`'' pair by
$\sigma _{i} $. The set of values $P_{i,i + 1}$, $N_{i}$, $\mu
_{i,i + 1}$, and $\sigma _{i}$ determines the force configuration
of the system. The geometrical configuration of the system is
represented by the angles between normal to the surface of contact
of two grains and the vertical as $\alpha _{i,i + 1} $. The force
equilibrium for each grain can be written as
\begin{equation}
\label{eq2}
\begin{array}{l}
 N_{i} - P_{i - 1,i} \vartheta_{i-1,i} - P_{i,i + 1} \vartheta_{i,i + 1} = 0 \\
 - m_{i} g + \sigma _{i} N_{i} - P_{i - 1,i} \theta_{i-1,i} + P_{i,i + 1} \theta_{i,i + 1} = 0 \,.
 \end{array}
\end{equation}
where
\[\begin{array}{ll}
\vartheta_{i,i + 1} = \sin \alpha_{i,i + 1} - \mu_{i,i + 1} \cos
\alpha_{i,i + 1} \,,\\
 \theta_{i,i + 1} = \cos \alpha _{i,i + 1} +
\mu _{i,i + 1} \sin \alpha _{i,i + 1}
\end{array}\]

We omit the equation for the moment of force to illustrate how
$\mu$-approach works. The 1D nature of the system permits to write
the expression for the the force between two particles, namely:
\begin{equation}
\label{eq3} P_{i,i + 1} = {\frac{{g}}{{\vartheta_{i,i +
1}}}}{\sum\limits_{k = 0}^{i} {{\frac{{m_{k} T_{k+1,i}}} {{\eta
_{k,k + 1} + \sigma_{k}}}}}}  + P_{0} T_{0,i} {\frac{{\vartheta_{
- 1,0}}} {{\vartheta_{i,i + 1} }}} \,,
\end{equation}
where \[T_{i_{1},i_{2}} = {\prod\limits_{n = i_{1}}^{i_{2}}
{{\frac{{\eta_{n - 1,n} - \sigma _{n}}} {{\eta_{n,n + 1} +
\sigma_{n} }}}}}\] and \[\eta_{i,i + 1} = {\frac{{1 + \mu_{i,i +
1} tg\alpha_{i,i + 1}}}{{tg\alpha_{i,i + 1} - \mu_{i,i + 1}}} }\]
and $P_{0}$ is the overweight.

Equation~(\ref{eq3}) allows to analyze the load distribution
through the silo with and without overweight with some
distribution of the contact friction coefficients $\mu_{i,i + 1}$.
\section{Results of the $\mu$-model for simple 1D packing}\label{section2}
It is well known that force distribution in granular media depend
not only on the silo height but also on the history and method of
preparation of the granular matter sample
\cite{vanel/pre/1999,howell99c}.
Within the approach proposed such a method can be modelled, e.g.
by the distribution of the coefficients $\sigma_{i}$, which shows
the stresses at the walls. Basing on the Eq. (\ref{eq3}) as the
exact microscopic solution of the model problem, we can model such
a feature, by the set of coefficients $\{\mu_{i,i + 1},\sigma
_{i}\}$ and angles $\{\alpha _{i,i + 1}\}$. These data determine
the geometrical and stress configuration of the system.

The question about deviation of static pressure distribution in
granular media from the Pascal law is of particular interest since
it is the characteristic feature of such kind of materials. Note
that that Eq.~(\ref{eq3}) reduces to the Pascal law with the
pressure being proportional to the depth if the walls are
absolutely smooth, i.e. $\sigma _{i} = 0$. This limiting case is
in correspondence with the fact that the deviation from the PL is
due to nonlinear dependence of the tangential component of the
stress along the wall.

\begin{figure}[hbt!]
\centering
\includegraphics[scale=0.7]{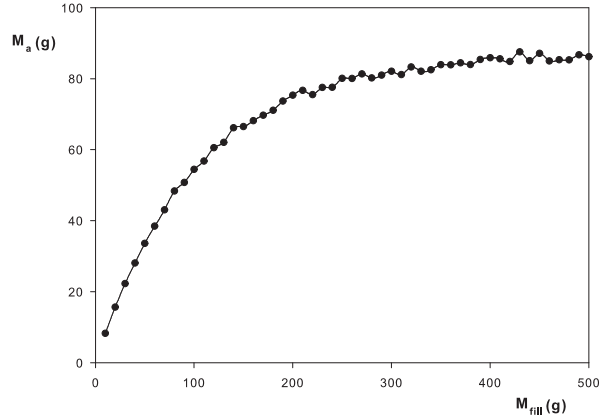}
\caption{Dependence the apparent mass on the filled mass of
Eq.~(\ref{eq3}) with random distributions of $\mu_{i,i+1}\in
[0.63,0.73]$ and $\sigma _{i}\in [0.25,0.27]$, and
$\alpha_{i,i+1}=\pi /4$, $m_{i}=10 g $. Data are comparable with
results of Ref. \cite{clement/epj/2005}.} \label{fig2}
\end{figure}

Other configurations give the deviation form the PL. We
investigate some of them numerically.
To compare our results with known experimental data we take the
random distributions of the friction coefficients $\sigma_{i}~\in~
[0.25, 0.27]$ and $\mu_{i,i + 1}\in [0.63, 0.73]$ with the
limiting values chosen as the best fits to the simulational values
of \cite{clement/epj/2005}, and $\alpha _{i,i + 1} = \pi / 4$.
Statistical averaging was performed over the 100 configurations
for 50 grains, each one of mass 10 g. The resulting dependence
between apparent and filled mass is shown in Fig.~\ref{fig2}.

\subsection{Force distribution in the silo without an overweight
in continuous limit}\label{sec2/1}

Another relatively simple case is the "homogeneous" granular
packing with $\mu_{i,i+1}=\mu$, $\sigma_{i}=\mu_{w}$ and
$\alpha_{i,i+1}=\alpha$. In this case one can take the continuous
limit in Eq.~\eqref{eq3}. For a silo without an overweight
($P_0=0$), one gets
\begin{equation}
\label{ph} P(\tilde {h}) = \rho g\lambda \left( {1 - e^{ - \tilde
{h} / \lambda}} \right)\,,
\end{equation}
with characteristic length
\begin{equation}
\label{lambda} \lambda = {\frac{{d}}{{\vartheta \cdot(\eta + \mu
_{w} ) \cdot \zeta \cdot \ln \left( {{\frac{{\eta + \mu _{w}}}
{{\eta - \mu _{w}}} }} \right)}}}
\end{equation}
where $\tilde{h}$ is the effective height:  $\tilde {h} = h \cdot
\vartheta \cdot (\mu _{w} + \eta )$, and configuration parameters
\[ \eta = {\frac{{1 + \mu \cdot \tan\alpha}}
{{\tan\alpha - \mu}} }\,,\quad \vartheta = \sin \alpha - \mu \cos
\alpha \,,\quad \zeta = {\frac{{1 + \sin \alpha}}{{\cos \alpha}}}
\] and $d$ is chute size.

So, we can see how the Janssen result can be obtained due to
microscopic approach.

Note, that if $\tan\alpha = \mu $, Eq.~\eqref{ph} again reduces to
the PL since grains do not lean against the wall. Another words
the static friction between grains is enough to keep them at rest
without any wall.
\begin{equation}\label{ph1}
P(h) = \rho gh \cdot \cos \alpha\,\,.
\end{equation}

Our result can be compared with Janssen formula \eqref{janss}.
Formula \eqref{janss} was obtained for vertical cylindrical chute.
To compare it with results of our model it should be obtained for
parallelepiped chute, which is infinite in one direction and has
the profile as shown on Fig. \ref{fig1}. Calculation for this case
transforms formula \eqref{janss} to the following:
\begin{equation}\label{myjanss}
p(z) = p_0 \left(1-\exp(-z/\lambda)\right)\,,\quad \lambda =
\f{d}{2 \mu k_J}
\end{equation}

\begin{figure}
\centering
\includegraphics[scale=0.7]{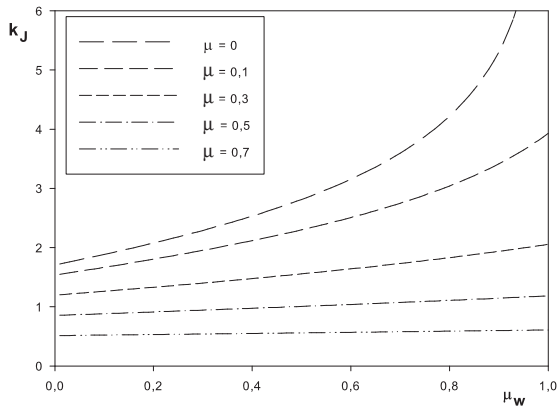}
\caption{The Janssen coefficient $k_{J}$ as a function of friction
$\mu _{w} $ on the walls from Eq.~\eqref{kj} for different $\mu
$.} \label{fig3}
\end{figure}

In the "homogeneous" state the analogue of the Janssen coefficient
$k_{J}$ can be devised via comparison of Eqs.~\eqref{ph} and
\eqref{lambda} with the result \eqref{myjanss}:
\begin{equation}
\label{kj} k_{J} = {\frac{{\vartheta \cdot \zeta}}{{2}}} \cdot
(\eta / \mu _{w} + 1) \cdot \ln \left( {{\frac{{\eta / \mu _{w} +
1}}{{\eta / \mu _{w} - 1}}}} \right).
\end{equation}
Equation (\ref{kj}) relates the Janssen coefficient with the
``microscopic'' characteristics of the granular packing. In Fig.
\ref{fig3} we illustrate the dependence of $k_{J} $ on friction at
the wall ($\mu _{w}$) for different values of internal friction
$\mu$.
\subsection{Force distribution in the silo with overweight}\label{sec2/2}
The influence of the $\mu$ distribution on the distribution of the
pressure can also be shown by considering the system when
overweight is present.

\begin{figure}[hbt!]
\centering
\includegraphics[scale=0.7]{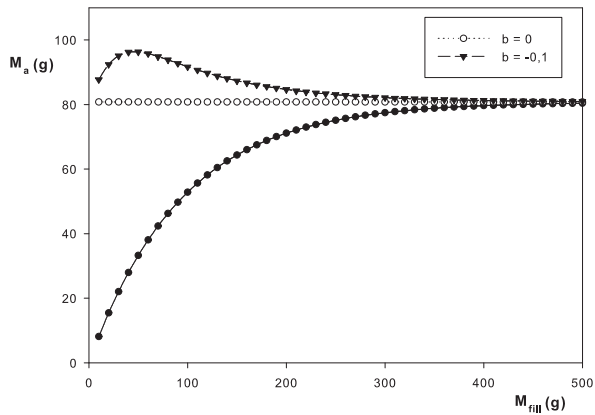}
\caption{Dependence of the apparent mass on the filled mass.
Bottom curve: force distribution without overweight. Middle curve:
force distribution with 80.8 g overweight for
$\mu_{i,\,i+1}=0.65$. Top curve: force distribution with 80.8 g
overweight for $\mu_{i,\,i+1}=0.65-0.1 e^{-\,0.5\, i}$. For all
curves $\sigma_{i} = 0.25$, \,$\alpha _{i,i + 1} = \pi / 4$,
$m_{i} = 10(g)$. Data are comparable with results of Ref.
\cite{clement/epj/2005}. }\label{fig4}
\end{figure}

For our numerical studies of the force distribution in the silo
with overweight we use $\sigma_{i}=0.25$ and value of overweight
$P_{0}=80.8 g$, which correspond to the value of grain-wall
friction coefficient and overweight of Ref.
\cite{clement/epj/2005}. Vessel contains 50 grains, each one of
mass 10 g. The resulting dependence between apparent and filled
mass is shown in Fig.~\ref{fig4}.
Bottom curve shows the force distribution in "homogeneous" media,
where all $\mu_{i,\,i+1}=\mu_{0}$ without overweight. As one  can
see from previous section this curve is nothing but Janssen
exponential distribution. In case of overweight presence in such a
"homogeneous" media, the force distribution is described by the
middle curve. This result was also predicted by Janssen but it
contradicts with the experiment. Since the force configuration is
determined by the distribution of the coefficients
$\mu_{i,\,i+1}$, we suppose that the latter has the same
functional character as that for the pressure without overweight.
The grounds of this assumption will be expanded in the next
section.

Choosing the $\mu$-distribution as following:
\begin{equation}\label{mui}
\mu_{i,\,i+1} = \mu_0 + b e^{-\,c\, i}\,,
\end{equation}
we adjust parameters $b,c$ so that to achieve the best fit (the
top curve on Fig.~\ref{fig4}) to data of
Ref.~\cite{clement/epj/2005}. Thus we can see, that the maximum in
the force distribution in the silo with overweight can be obtained
if the friction coefficients are changed in the same way as the
pressure changes without an overweight.
\section{Discussion of the $\mu$-model}\label{section3}

The results obtained in previous section for simplified 1D model
within the framework of $\mu$-model can be extended further since
the final results do not contain any microscopic characteristics
of such a specific model. Indeed, the proposed approach allows us
to switch the description of the granular media from the
consideration of force network to the distribution of friction
coefficients, which in continuous limit transforms to the field of
tensor $\mu$. Such a field becomes an additional characteristic of
granular media which cannot be obtained due to common approach
(e.g from Newton's equations). Additional statistical arguments
about how this field is distributed should be used.

As one can see from Sec.~\ref{section2}, we modelled the
distribution of the $\mu$ in two different ways. First was uniform
distribution $\mu=const$, and we showed how such an assumption
conforms with previous theoretical results and experimental data.
Another distribution was of exponential form and here we give the
grounds for such a choice.

\subsection{Spatial distribution of the $\mu$-field}\label{sec3/1}
Let us consider the distribution of pressure in a silo with
additional weight on its top. Suppose, that overweight is
implemented by another silo of the same material, in which the
pressure has reached its saturated value. Thus, the considered
part of the granular media is actually belongs to the region where
the pressure has reached it's saturated value. Therefore the
pressure in this part must be equal to overweight. These
reasonings are confirmed by Janssen's model \cite{janssen} and our
results, in which it is assumed that friction either at the wall
or in the bulk is $const$. This is in obvious contradiction with
the experiment \cite{ovralezclement/pre/2003}, which shows that
there is a maximum in the pressure dependence on depth.

It is possible to get such a nonmonotonic dependence of pressure
on overweight in some theoretical approaches (see e.g.
\cite{pitman/pre/1998}), but there is no clear explanation why it
appears. In addition, they are based on the assumptions $\mu =
const$, which is adequate only near the Coulomb threshold. As we
can see from the experiment, there are the "screening" region of
size $\lambda$. In case of absence of the overweight, the pressure
increases there and in case of overweight presence, the pressure
has maximum within this region. Under overloading the stress
configuration within this region changes more drastically than in
the bulk. Within the approach proposed it can be described as the
spatial distribution of the coefficients $\mu_{i,i+1}$ and
$\sigma_{i}$, in general. In continuum approach it corresponds to
some dependence $\mu(z)$ in the bulk and $\sigma(z)$ at the
boundary.

Since $\mu$ characterizes the stress configuration, its spatial
character must be similar to that of pressure in the system
without overweight. This assumption can be viewed as a first
approximation in the expansion spatially distributed function in a
series of approximation. Indeed the function $\mu(z)$ can be
written as
\begin{equation}\label{mu0}
\mu(z) = \mu_0 + \mu_1(z) + \mu_2(z) + \ldots\,\,.
\end{equation}
Here $\mu_0$ is constant part of $\mu$, which is used in most of
theoretical descriptions. Within the approach proposed $\mu_1(z)$
changes in the same way as the pressure without overweight
changes. Then $\mu_2$ takes into account changing in pressure in
granular medium where $\mu(z)= \mu_0 + \mu_1(z)$, etc.

\begin{figure}
\centering
\includegraphics[scale=0.7]{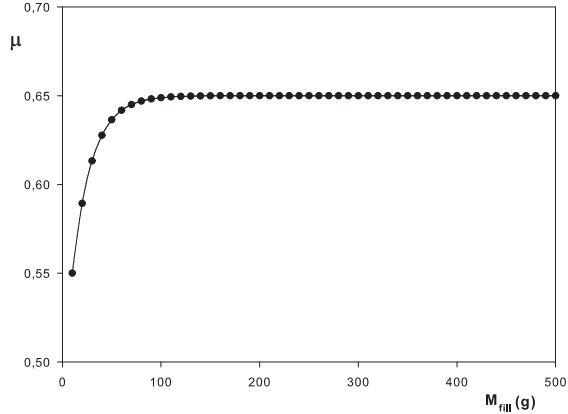}
\caption{Dependence of the spatially distributed $\mu$ on the
depth in units of filled mass.}\label{fig5}
\end{figure}

As one  can see from previous section, to satisfy the experimental
results, it is enough to model the distribution of $\mu_{i,i+1}$
as:
\begin{equation}\label{mui1}
\mu_{i,i+1} = \mu_0 + b e^{-c i}\,,
\end{equation}
omitting higher order approximations. We take $\mu_0$ as the bulk
value, coefficient $b$ itself governs the gap between bulk and
border value of $\mu$, both $b$ and $c$ are responsible for the
speed of $\mu$ increasing.
At the values of parameters, found in Sec.~\ref{section2}, $\mu$
becomes saturated very quickly, so we have very thin region where
$\mu$ changes from $0.55$ to $0.65$ (see Fig.~\ref{fig5}).

But, as one can see from Fig~\ref{fig4}, this region, where $\mu$
has such a dependence is enough to change the pressure
distribution from the flat curve to the curve with the maximum.
There is no maximum in the pressure dependence with depth, when
$\mu=const$. This result, which was predicted by Janssen model, is
confirmed by physical arguments, given above. Maximum appears when
$\mu$ is distributed within a bulk in a given way. To satisfy the
experimental results, region where $\mu$ changes in such a way
must be narrow. Thus we can say about changing only near-boundary
values of $\mu$.

Note, that we have an essential difference in pressure
dependencies only if overweight is present. When there is no
overweight, the pressure dependence for constant $\mu$ and the
pressure dependence for changing $\mu$ are almost the same. We can
see that presence of an overweight reveals the real distribution
of $\mu$.

So we can conclude the following: a) the value of $\mu$ is not
constant and has some distribution; b) such a  distribution can be
revealed only by the presence of an overweight and this
distribution is essential only near the boundary.

The last fact can be explained in the following way. When we fill
the silo with the granular material some stress configuration is
implemented in it. If then we put an overweight at the top of the
silo, that configuration will become broken, and new one will be
implemented. This happens because near-boundary layer feels this
overweight and react on it. Other layers do not feel the
overweight in essential way because of jamming in the upper layer.
Other words, to fill the silo to height $2h$ at once is not the
same that to fill the silo to height $h$ first and then $h$ again.
This also illustrates how the stress distribution in the granular
media depends on the history of packing creation.

\subsection{Macroscopic parameters}\label{sec3/2}
\begin{figure}[hbt]
\centering
\includegraphics[scale=0.7]{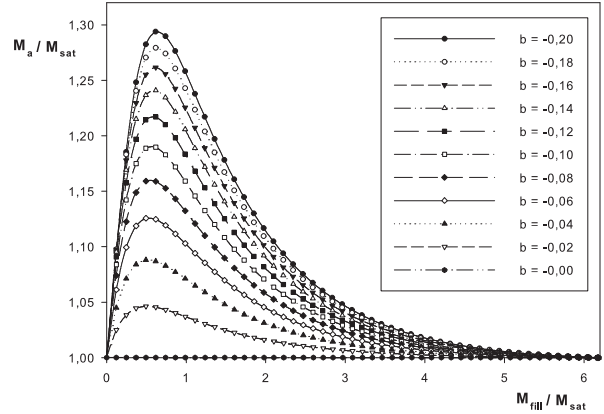}
\caption{Rescaled dependencies the apparent mass on the filled
mass for different parameters $b$, which changes from $-0.2$ to
$0$ through $0.02$. The bottom curve corresponds to $b=0$, the
upper one corresponds to $b=-0.2$. Overweight and saturated
pressure equals $80.32$ g. Apparent and filled mass expressed in
overweight units. Other parameters are the same as for
Fig.~\ref{fig4}.}\label{fig6}
\end{figure}

Previous consideration allows to conclude that there must be at
least two characteristic scales which characterize the state of
silo under overloading. Followed by \cite{clement/epj/2005} we
built rescaled dependencies (see Fig.~\ref{fig6}).

\begin{figure}
\centering
\includegraphics[scale=0.7]{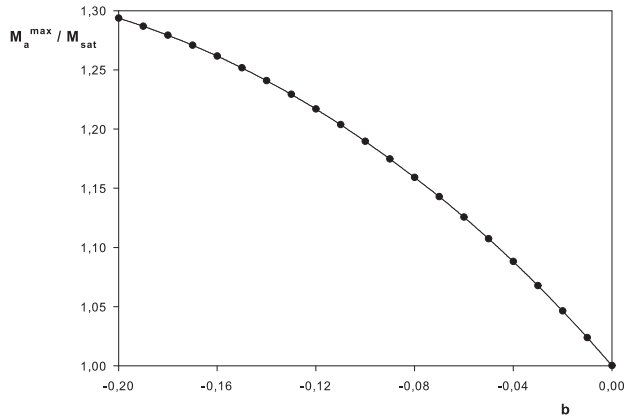}
\caption{Dependence the maximum of apparent mass on the $\mu$
changing gap in case of 80.83 g overweight presence.}\label{fig7}
\end{figure}

As one can see the bottom curve corresponds to the $\mu=const$
because the fact that $b=0$ is equivalent to the
$\mu_{i,i+1}=\mu_0$. For other values of $b$ $\mu$ is not $const$
and thus the apparent mass is not $const$ either. The distribution
of the apparent mass on filled mass has maximum, with value which
depends on $b$ shown at Fig.~\ref{fig7}.

One can see from this that there is no universal rescaled curve
for different parameters. This fact was also mentioned in
\cite{clement/epj/2005}. Within the approach proposed this is the
consequence of the existence of  additional scale parameter
$\lambda_\mu$. Such a scale becomes evident in the presence of
overweight if $\mu$ is not constant but $\mu$ is distributed
within bulk. According to the proposed distribution of the $\mu$,
$\lambda_\mu$ depends on $b$ and $c$, the gap in which $\mu$
changes and the steepness of $\mu$ saturation.

The dependencies shown on Fig.~\ref{fig6} must be parameterized
using this parameter. In other words, this parameter splits
Janssen's curve if silo is overloaded.

\section{Conclusion.}\label{conclusion}
We propose the description of the granular media at rest based on
the introduction of the spatial distribution for contact
coefficients $\mu$ of dry friction. With the help of the simple
static model we investigate the distribution of the weight in the
silo. It is shown that in the case without overweight and
homogeneous distribution of the friction coefficients the Janssen
result is recovered. In a case of overweight we predict the
maximum for the apparent mass as a function of a filled one, which
is observed in experiments. It is important that the nature of
such a maximum is related to the inhomogeneity in the spatial
distribution of the dry friction coefficients. Such a distribution
is formed due to the jamming of the grains in the upper layers
which bear most of the overload. We put forward physical arguments
which allow to obtain such distribution of the friction
coefficients by taking into account the inhomogeneity of the
pressure distribution with the height. Note the these result are
obtained without any assumption about elasticity of the grains
made in quasielastic approaches \cite{clement/epj/2005}. In
addition within the proposed approach it is possible to explain
the absence of simple rescaling law for the overshooting effect by
the presence of two characteristic length. The first length is the
Janssen length $\lambda_J$ and characterizes the distribution of
pressure, the other one is the scale of spatial distribution for
dry friction coefficient.

In the continuous limit the development of the approach proposed
implies the introduction of the tensor field for dry friction.
Such a field becomes additional characteristic of a granular
medium which is determined by both the geometrical and the stress
configurations. The possibility of such a description is due to
the fact that the interparticle interactions in the granular media
are of short distance character \cite{ll7}. It gives the grounds
to expect that at least the static of granular medium should be
described with the traditional framework of general elasticity
theory with proper modifications.

\begin{acknowledgments}
Authors thank to Dr. K.~Shundyak for numerous discussions about the
results. His critical remarks were also substantial for the final
form of the paper.
\end{acknowledgments}

\end{document}